\documentclass[conference]{IEEEtran}
\IEEEoverridecommandlockouts
\usepackage{cite}
\usepackage{amsmath,amssymb,amsfonts,amsthm}
\usepackage{graphicx}
\usepackage{subcaption}
\usepackage{xcolor}
\usepackage{color,soul}
\usepackage{bbm}
\usepackage{comment}
\usepackage{textcomp}
\usepackage{algorithm,float,algpseudocode}
\usepackage{lipsum}
\usepackage[utf8]{inputenc}
\usepackage[english]{babel}

\newcounter{procedure}
\makeatletter

\newtheorem{theorem}{Theorem}

\soulregister\cite7
\soulregister\ref7
\soulregister\eqref7
\soulregister\pageref7



\begin{document}

\setlength{\textfloatsep}{2pt}
\setlength{\floatsep}{2pt}
\setlength{\parskip}{2pt}
\setlength{\abovedisplayskip}{2pt}
\setlength{\belowdisplayskip}{1.5pt}
\setlength{\abovecaptionskip}{2pt}

\title{Age-Optimal Power Allocation in Industrial IoT: A Risk-Sensitive Federated Learning Approach}
\author{\IEEEauthorblockN{Yung-Lin Hsu\IEEEauthorrefmark{1}, Chen-Feng Liu\IEEEauthorrefmark{2}, Sumudu Samarakoon\IEEEauthorrefmark{2}, Hung-Yu Wei\IEEEauthorrefmark{1}, and Mehdi Bennis\IEEEauthorrefmark{2}}
\IEEEauthorblockA{\IEEEauthorrefmark{1}Graduate Institute of Communication Engineering, National Taiwan University, Taiwan\\\IEEEauthorrefmark{2}Centre for Wireless Communications, University of Oulu, Finland}
E-mail: \{d04942010, hywei\}@ntu.edu.tw, \{chen-feng.liu, sumudu.samarakoon, mehdi.bennis\}@oulu.fi}

\maketitle

\begin{abstract}
This work studies a real-time environment monitoring scenario in the industrial Internet of things, where wireless sensors proactively collect environmental data and transmit it to the controller. We adopt the notion of \emph{risk-sensitivity} in financial mathematics as the objective to jointly minimize the mean, variance, and other higher-order statistics of the network energy consumption subject to the constraints on the age of information (AoI) threshold violation probability and the AoI exceedances over a pre-defined threshold. We characterize the extreme AoI staleness using results in \emph{extreme value theory} and propose a distributed power allocation approach by weaving in together principles of Lyapunov optimization and \emph{federated learning} (FL). Simulation results demonstrate that the proposed FL-based distributed solution is on par with the centralized baseline while consuming $28.50\%$ less system energy and outperforms the other baselines.
\end{abstract}

\vspace{-.7em}
\section{Introduction}
Environment monitoring and control in smart factory scenarios are mission-critical applications in 5G and beyond, where sensors, meters, and monitors generate and upload data to a central controller with real-time ultra-low latency. 
In particular, for real-time monitoring and control, the elapsed time since the data was generated is a key factor for the control performance.
Such time duration is referred to as the \emph{age of information} (AoI). 
If the AoI of the controller's available data grows unexpectedly, the outcome of the real-time environmental monitoring and control will be poorly degraded \cite{zhou2020age}. 
The impact of AoI-aware resource allocation has been investigated in various communication systems \cite{zhou2020age,Liu_2019,8904450,8766741,8969641}.
In \cite{zhou2020age}, a mean-field game approach was proposed in a dense IoT monitoring system. 
The work \cite{Liu_2019} considered a multi-sensor industrial IoT (IIoT) scenario with finite blocklength transmission in which the controller instructs all devices to sample and upload environment data based on its AoI records.
The objective therein was to minimize the sensors' power consumption.
Aiming at minimizing the average AoI and average peak AoI using a Markov decision process.
The work \cite{8904450} investigated the tradeoff between AoI and energy cost and proposed an action policy for devices. 
In \cite{8766741}, the authors assumed that the devices could not upload data while wirelessly harvesting energy from the base station. 
As a result, the AoI exponentially increases during the energy harvesting period. 
Finally, the authors in \cite{8969641} considered a remote monitoring problem trading off the expected AoI and the AoI threshold violation probability.

While reducing the AoI in a centralized manner, the proposed resource allocation approaches in \cite{zhou2020age,Liu_2019,8904450,8766741,8969641} incurred not only supernumerary power for statistical information exchanging but also tremendous signaling overheads, which are not negligible in ultra-low latency real-time monitoring, e.g., industrial automation. 
Under the ultra-low latency constraints, delegating the transmission decisions to the sensors provides a more realistic avenue, especially when the data-sampling time is uncertain. 
Therefore, while accounting for the AoI threshold violation and the threshold-exceeding events, our work aims at proposing a decentralized and proactive power allocation scheme by minimizing the sensors' entropic risk measure energy consumption \cite{8936848}.
%Minimizing the sensors' energy consumption variance prevents some of the sensors from consuming superabundant energy.
We further leverage \textit{extreme value theory} at the sensor to characterize the AoI exceedances over a threshold using the locally observed historical data. 
However, the accuracy of characterization is limited by the amount of the sensor's data.
To improve the accuracy under limited data availability, we resort to \emph{federated learning} (FL),  a collaborative and distributed model training framework \cite{park2019wireless,park2020communicationefficient}, and propose an FL-based distributed power allocation algorithm along with Lyapunov optimization. Numerical results show that the proposed approach is on par with the centralized baseline while consuming less system energy for data transmission and model training.

\vspace{-.4em}
\section{System Model and Problem Formulation\vspace{-0.3em}}
\begin{figure}[t]
	\begin{center}
	\includegraphics[width = \columnwidth]{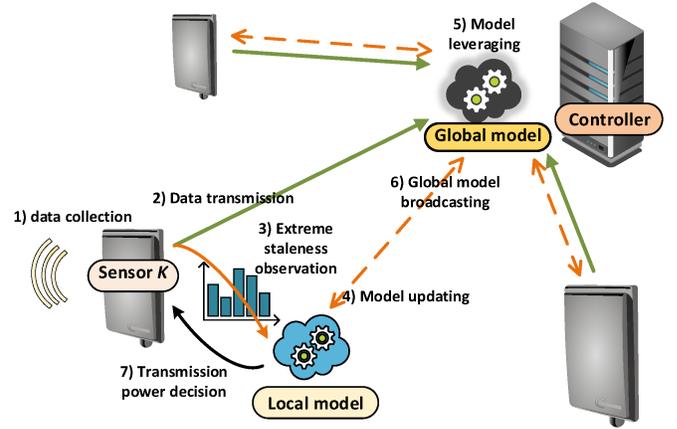}
	\caption{Sensor-Controller system architecture.}
	\label{fig:simple_scenario}
	\end{center}
\end{figure}
\begin{figure}[t]
	\begin{center}
	\includegraphics[width = \columnwidth]{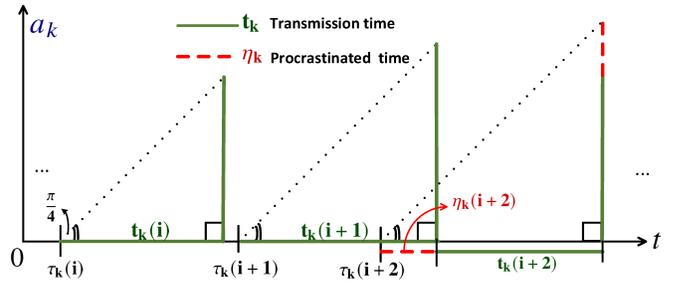}
	\caption{The AoI instances with data sampling and transmission. \small Since only $\eta_k(i+2)>0$, we have $a_k(i)=t_k(i)$, $a_k(i+1)=t_k(i+1)$, and $a_k(i+2)=\eta_k(i+2)+t_k(i+2)$.}
	\label{fig:AoI_example}
	\end{center}
\end{figure}

 
As shown in Fig.~\ref{fig:simple_scenario}, we consider an industrial IoT network with a set $\mathcal{K}$ of $K$ intelligent sensors that monitor distinct and independent (but of the same type) environments and transmit the sampled data to a controller. 
%The sensors cannot communicate with one another due to their geographical locations.
We assume that the sensors' sampling operations are event-triggered by the environmental changes without any available statistical information.
%, and the triggering time is random without any available statistical information. 
After the sensor samples the status data, the new data is immediately transmitted to the controller if the previous data's uploading procedure has been completed. Otherwise, the (new) data is queued in the buffer. 
Such a queueing time is called the procrastinated time in this paper. 
%If the uploading procedure (of the previous data) has not been completed, the (new) data is queued in the buffer. 
Let us index the sequentially sampled data of each sensor by $i\in\mathbb{Z}^{+}$ and denote the procrastinated time of the $i$th data of sensor $k\in\mathcal{K}$ as $\eta_k(i)\geq 0$.
Then to send the $i$th data with size $N$, sensor $k$ allocates transmission power $p_k(i)$ over its dedicated bandwidth $B$ with the transmission time
\begin{equation}\label{Tx_time}
    t_k(i) = \frac{N}{B\log_2\big(1+\frac{h_k(i)p_k(i)}{N_0 B}\big)},
\end{equation}
in which $h_k(i)$ is the channel gain, including path loss and channel fading, between sensor $k$ and the controller, and $N_0$ is the power spectral density of the additive white Gaussian noise. Further, assume that the controller issues control commands to the actuator right after receiving the status data. Hence, we focus on the AoI of the time instant at which the status data is fully received at the controller. In this regard, we express the concerned AoI of the $i$th data as $a_k(i) = \eta_k(i) + t_k(i)$ which incorporates the elapsed procrastinated and transmission time.
The data sampling and transmitting instances are schematically illustrated in Fig.~\ref{fig:AoI_example}. Therein, $\tau_k(i)$ denotes the sampling time instant of the $i$th data, and the procrastinated time is shown as the overlapping period.
From Fig.~\ref{fig:AoI_example}, we can straightforwardly find $\eta_k(i)=[\tau_k(i-1) + a_k(i-1) - \tau_k(i)]^{+}$ with $[x]^+=\max\{x,0\}$ and rewrite the AoI in a recursive manner as
\begin{equation}\label{AoI}
    a_k(i) = [\tau_k(i-1) + a_k(i-1) - \tau_k(i)]^{+}+ t_k(i).
\end{equation}
%
%
%
%where $[x]^+=\max\{x,0\}$. 
Since the information ages in consecutive transmissions are entangled as per \eqref{AoI}, we should proactively account for the impacts of the allocated power on future AoI.
%The data sampling and uploading instances are schematically illustrated in Fig.~\ref{fig:AoI_example}. 

Note that the factory environment varies continuously, in which the stale information can degrade the control system performance. 
As a remedy, we impose a set of constraints on the AoI in terms of the weighted expectation, AoI outage probability, and AoI exceedance.
We firstly introduce a staleness function from \cite{staleness_function} as
%
%
%
%\begin{equation}\label{stalemess_function}
%    f_k(i) = \frac{[a_k(i)]^{(1-\beta)}}{1-\beta}
%\end{equation}
$f_k(i) = \frac{[a_k(i)]^{(1-\beta)}}{1-\beta}$
for the AoI staleness with a predetermined value $\beta \leq 0$ and then consider a long-term time-averaged constraint for every sensor $k$, i.e.,
\begin{equation}\label{AoI_loss_bound}
    \lim\limits_{I \to \infty} \frac{1}{I}\sum^I_{i=1} \mathbbm{E}[f_k(i)] \leq f_0,\forall k \in \mathcal{K},
\end{equation}
where $f_0$ is a pre-defined threshold. Additionally, we impose a probabilistic constraint on the AoI for each sensor $k$ as
\begin{equation}\label{outg}
 \lim\limits_{I \to \infty} \frac{1}{I}\sum^I_{i=1}\Pr\{f_k(i)>f_0\} \leq \epsilon, \forall  k \in \mathcal{K},
\end{equation}
in which $\epsilon\ll 1$ is the tolerable threshold violation probability.
Although a very low occurrence probability of the AoI exceedance is ensured in \eqref{outg}, the uploaded data with an extremely large age can hinder the control performance. 
To mitigate this effect, a constraint on the AoI exceedances $q_k(i) = f_k(i)-f_0>0$ is imposed as follows:
\begin{equation}\label{extreme_AoI}
   \lim\limits_{I \to \infty} \frac{1}{I}\sum^I_{i=1} \mathbbm{E}[q_k(i)] \leq e_0, \forall k \in \mathcal{K}.
\end{equation}
Here, we define the set of AoI exceedance of sensor $k$ as $\mathcal{Q}_k \triangleq \{q_k(i)|\mathbbm{1}_{\{q_k(i)>0\}}, \forall i\}$, and $e_0$ is a pre-defined threshold.
Finally, taking into account the sensors' limited-energy, we aim at minimizing not only the sensors' average energy consumption but also the variance for data uploading to reduce superabundant energy between each sensor. 
To this end, we invoke  the entropic risk measure $\frac{1}{\rho}\ln(\mathbbm{E}_{X}[\exp({\rho X})])$,
which incorporates the mean, variance, and higher-order statistics of $X$ \cite{8936848}, in our global objective, and formulate the studied optimization problem as follows:
\begin{subequations}\label{origin}
 \begin{IEEEeqnarray}{cl}
 \mathop{\text{minimize}}\limits_{p_k(i)}
 &~~ \frac{1}{\rho}\ln\bigg( \lim\limits_{I \to \infty} \frac{1}{I}\sum^I_{i=1} \frac{1}{K}\sum_{ k\in\mathcal{K}}\exp({\rho E_k(i)})\bigg)\label{main_problem}\\
 \text{subject to}
 &~~0\leq p_k(i) \leq p_{max}, \forall i \in \mathbb{Z}^+, k \in \mathcal{K}, \label{PW_max}\\
 &~~ \eqref{AoI_loss_bound}, \eqref{outg},\mbox{and\,} \eqref{extreme_AoI}, \nonumber
 \end{IEEEeqnarray}
 \end{subequations}
 where $E_k(i)=p_k(i)t_k(i)$ denotes sensor $k$'s energy consumption for transmitting its $i$th data,
 $\rho>0$ reflects the weights of the variance and higher-order statistics in risk minimization, and $p_{max}$ is the sensor's data transmission power budget.
Moreover, in the objective \eqref{main_problem}, we take the average over time and sensors instead of expectation in the entropic risk measure.
In the next section,  we will utilize Lyapunov optimization techniques  \cite{Lyopunov} to solve problem \eqref{origin} since the objective \eqref{main_problem} and constraints \eqref{AoI_loss_bound}, \eqref{outg}, and \eqref{extreme_AoI} are long-term time-averaged functions.

\section{FL-Based Distributed Power Allocation}\label{proposed_scheme}
\subsection{Lyapunov Optimization Framework}
In order to ensure the time-averaged constraints by Lyapunov optimization, we first introduce the virtual queues 
\begin{align}
    \Gamma_k(i+1)&=[\Gamma_k(i)+f_k(i)-f_0]^+,\label{Virtual_queue_1}
\\    \Upsilon_k(i+1)&=[\Upsilon_k(i)+[q_k(i)-e_0]\mathbbm{1}_{\{q_k(i)>0\}}]^+,\label{Virtual_queue_2}
\end{align}
 for constraints \eqref{AoI_loss_bound} and \eqref{extreme_AoI}, respectively. Additionally, by applying $\Pr\{f_k(i)>f_0\}=\mathbb{E}[\mathbbm{1}_{\{q_k(i)>0\}}]$ and scaling both sides of \eqref{outg} as $\lim\limits_{I \to \infty}\frac{1}{I}\sum_{i=1}^I f_k(i) \mathbb{E}[\mathbbm{1}_{\{q_k(i)>0\}}] \leq \lim\limits_{I \to \infty}\frac{1}{I}\sum_{i=1}^I f_k(i)\epsilon$, the virtual queue
\begin{equation}\label{Virtual queue 3}
    \Lambda_k(i+1)=[\Lambda_k(i)+(\mathbbm{1}_{\{q_k(i)>0\}}-\epsilon)f_k(i)]^+
\end{equation}
is considered for the constraint \eqref{outg}.
Due to the fact that minimizing $\lim\limits_{I \to \infty} \frac{1}{IK}\sum^I_{i=1}\sum_{ k\in\mathcal{K}}\exp(\rho E_k(i))$ is equivalent to minimizing \eqref{main_problem}, we consider the conditional Lyapunov drift-plus-penalty
%Subsequently, applying $([Q+y]^+)^2\leq Q^2+2Qy+y^2$ and all queue length evaluations to the conditional Lyapunov drift-plus-penalty function \cite{Lyopunov} with the fact that \eqref{main_problem} is equivalent to minimizing $ \lim\limits_{I \to \infty} \frac{1}{I}\sum^I_{i=1} \mathbbm{E}[e^{\rho E_k(i)}]$, i.e.,
%
%
%
\begin{multline}\label{Lyapunov}
\frac{1}{K}\sum_{k \in \mathcal{K}}\mathbb{E}\Big[\frac{1}{2}\boldsymbol{\Phi}_k(i+1)\boldsymbol{\Phi}_k(i+1)^{\rm T}-\frac{1}{2}\boldsymbol{\Phi}_k(i)\boldsymbol{\Phi}_k(i)^{\rm T}\\+V\exp(\rho E_k(i))\Big],
\end{multline}
where $\boldsymbol{\Phi}_k(i)=[\Gamma_k(i),\Upsilon_k(i),\Lambda_k(i),\forall k \in \mathcal{K}]$ is the combined queue vector for notational simplicity, and $(\cdot)^{\rm T}$ represents the transpose of a vector.  Subsequently, by applying $([Q+y]^+)^2\leq Q^2+2Qy+y^2$, we can derive the inequality
\begin{equation}\label{L_d_bound}
\eqref{Lyapunov}\leq 
\frac{1}{K}\sum_{k \in \mathcal{K}}\mathbb{E}\big[\Delta_0+F_k(i)+V\exp(\rho E_k(i)) \big],
\end{equation}
in which
$  F_k(i) = \theta^1_k(i)[f_k(i)]^2 + \theta^2_k(i)f_k(i)$
with
$\theta^1_k(i)=\frac{1}{2}(1+\epsilon^2)+(1-\epsilon)\mathbbm{1}_{\{q_k(i)>0\}}$ and
$\theta^2_k(i)=\Gamma_k(i)-f_0-\epsilon\Lambda_k(i)
+[\Lambda_k(i)+\Upsilon_k(i)-f_0-e_0]\mathbbm{1}_{\{q_k(i)>0\}}$.
Additionally, $\Delta_0=\frac{1}{2}[f^2_0-\Gamma_k(i)+[(f_0+e_0)^2-\Lambda_k(i)(f_0+e_0)]\mathbbm{1}_{\{q_k(i)>0\}}$ is a constant.
%and $V > 0$ is the tradeoff parameter between the lengths of the virtual queues and the optimality of the energy consumption in \eqref{origin}.
Note that the solution of \eqref{origin} can be obtained by minimizing the derived upper bound on the conditional Lyapunov drift-plus-penalty function  \cite{Lyopunov}, i.e.,  \eqref{L_d_bound}, in each transmission $i$ by optimizing over the transmit power $p_k(i)$. 
Furthermore, $V \geq  0$ is the tradeoff parameter between the lengths of the virtual queues and the optimality of the energy consumption in \eqref{origin}.

\vspace{-0.4em}
\subsection{Distributed Power Allocation with Federated Learning for AoI Exceedances \vspace{-0.3em}}
Obtaining the optimal solution of minimizing the upper bound \eqref{L_d_bound} requires the global information about $\boldsymbol{\Phi}_k(i), \forall k$. 
However, centrally allocating power (i.e., the centralized model-training (CENT) scheme in Section \ref{Numerical}) consumes tremendous power overheads on queue state information exchanges.
To improve power efficiency, we propose a decentralized FL-Based scheme, wherein each sensor $k\in\mathcal{K}$ locally allocates its transmit power for each transmission $i\in\mathbb{Z^{+}}$ by minimizing the individual upper bound of \eqref{L_d_bound},
 \begin{align}\label{L_form}
 \underset {p_k(i)}{\text{minimize}}
 ~V\exp(\rho E_k(i)) +F_k(i)
 ~~\text{subject to}
 ~\eqref{PW_max}.
 \end{align}
In \eqref{L_form}, we can straightforwardly prove the convexity of $F_k(i)$, but $\exp(\rho E_k(i))$ is non-convex with respect to $p_k(i)$.
In order to tractably solve the non-convex problem \eqref{L_form}, we adopt the notion of the convex-concave procedure (CCP) \cite{CCP} by which we iteratively convexify the non-convex part $\exp(\rho E_k(i))$ by the first-order Taylor series expansion with respect to a reference point $\hat{p}_k$ as
\begin{equation*}
    \textstyle \exp(\rho E_k(\hat{p}_k))\left[t_k(\hat{p}_k )\left(1-\frac{\hat{p}_k h_k t_k(\hat{p}_k )}{ N(\hat{p}_k h_k+N_0 B)\ln2}\right)(p_k-\hat{p}_k)+1\right]
\end{equation*}
and solve the convexified problem. Specifically, given the reference point $\hat{p}_k^{r}$ in the $r$th iteration, we focus on
\begin{subequations}\label{CCP_r}
 \begin{IEEEeqnarray}{cl}
 \mathop{\text{minimize}} \limits_{\hat{p}_k^{r+1}}
 &~~ VJ(\hat{p}_k^{r})\hat{p}_k^{r+1} + F_k(\hat{p}_k^{r+1})\label{CCP_r_obj}\\
 \text{subject to}
 &~~ 0\leq \hat{p}_k^{r+1} \leq p_{max},
 \end{IEEEeqnarray}
\end{subequations}
in which $J(\hat{p}_k^{r})=\exp(\rho E_k(\hat{p}_k^{r}))t_k(\hat{p}_k^{r})\big[1-\frac{\hat{p}_k^{r} h_k t_k(\hat{p}_k^{r})}{N(\hat{p}_k^{r} h_k+N_0 B)\ln 2}\big]$.
The optimal solution to \eqref{CCP_r} is subsequently used as the reference point $\hat{p}_k^{r+1}$ in the next iteration.
The initial reference point $\hat{p}_k^0$ is randomly selected from $(0,p_{max}]$. When $|\hat{p}^{r}_k-\hat{p}^{r-1}_k|\to 0$ after a number of iterations, sensor $k$ uses the transmit power $p^*_k(i)=\hat{p}^{r}_k$ to upload the $i$th data and then updates \eqref{AoI} and all virtual queues \eqref{Virtual_queue_1}, \eqref{Virtual_queue_2}, and \eqref{Virtual queue 3}. 

However, since the AoI exceedance happens rarely, the deviation of virtual queue \eqref{Virtual_queue_2} between each sensor becomes severe with local statistics only.
To address this issue, we invoke results from extreme value theory and the principles of FL. 
Let us rewrite the virtual queue $\Upsilon_k(i)$ in \eqref{Virtual_queue_2} as
    \begin{align*}
        \textstyle\Upsilon_k(i+1)&=\sum^{i}_{j=1}[q_k(j)-e_0]^+\mathbbm{1}_{\{q_k(j)>0\}}
        \\
        & \textstyle \geq \left [ \sum^{i}_{j=1}[q_k(j)-e_0]\mathbbm{1}_{\{q_k(j)>0\}} \right ]^+
        \\
        &\textstyle \stackrel{(a)}{=}\left[ \frac{\sum^{i}_{j=1}q_k(j)\mathbbm{1}_{\{q_k(j)>0\}}}{\sum^{i}_{j=1}\mathbbm{1}_{\{q_k(j)>0\}}}-e_0 \right ]^+\sum^{i}_{j=1}\mathbbm{1}_{\{q_k(j)>0\}}
    \end{align*}
in which the first term in $(a)$ represents the empirical average, which may have large variance due to limited historical data of the excess AoI.
Nevertheless, if the mean of the AoI exceedance is available, we can estimate the steady-state average length of the virtual queue \eqref{Virtual_queue_2}.
\begin{theorem}[Pickands–Balkema–de Haan theorem\cite{Coles2001}] \label{thm1}
Given a random variable $A$ with the cumulative distribution function $F_A(a)$ and a threshold $a_0$, as $a_0 \to F^{-1}_A(1)$, the excess value $Q=A-a_0>0$ can be approximately characterized by a generalized Pareto distribution (GPD) with the scale  $\sigma>0$ and shape $\xi\in\mathbbm{R}$ parameters.
The mean of the GPD is $\frac{\sigma}{1-\xi}$.
\end{theorem}
Leveraging the results in Theorem \ref{thm1}, we characterize the statistics of $q_k(j)$ as a GPD whose parameters $\sigma$ and $\xi$ (i.e., the mean) can be estimated using maximum likelihood estimation.
Given a sufficient amount of historical data, the GPD model (i.e., scale and shape parameters) of the excess AoI can be trained.
However, owing to the sparsity of the excess AoI data at the sensor, it is time-consuming for each sensor to train the GPD model independently.
To overcome this hurdle, we utilize the FL framework in which
all sensors periodically update their locally-trained model to the controller. 
Then the controller aggregates the updated local models and feeds back the aggregated model to the sensors. 
Our FL-based model training is detailed as follows.

Assume that the local-model updating time interval is $M$, and each interval is indexed by $m \in \mathbb{Z}^+$. 
In every updating time interval, each sensor trains its model locally.
In order to have sufficient independent data for local training, we set $W$ observation time windows within which the sensor selects the largest excess AoI as a training sample. 
The observation time windows are indexed by $w\in\mathbb{Z}^{+}$, and the window size is $O$ with $M/O=W\in\mathbb{Z}^{+}$, which should be sufficiently large to minimize the correlation between the selected data while being sufficiently small to prevent filtering out the data overmuch. 
Moreover, the selected extreme data at sensor $k$ in the $w$th time window of the $m$th time interval is denoted by $\hat{q}^{m,w}_k=\max_{\tau_k(i) \in\mathcal{T}^{m,w}}\{q_k(i)|\mathbbm{1}_{\{q_k(i)>0\}}\}$, where $\mathcal{T}^{m,w}\in [M(m-1)+O(w-1),M(m-1)+Ow]$. 
The selected data set within the $m$th time interval is denoted by $\mathcal{Q}^m_k=\{\hat{q}^{m,w}_k\}^{W}_{w=1}$.
After collecting the samples $\mathcal{Q}^m_k$, we train the sensor $k$'s GPD model $\theta^m_k = \{\sigma^m_k, \xi^m_k\}$ via \textit{tilted empirical risk minimization} (TERM) \cite{li2020tilted}, i.e.,
\begin{multline}
 \underset{\theta^m_k}{\mbox{minimize}} ~   \tilde{\mathcal{L}}(\theta^m_k|t,\mathcal{Q}^m_k)
 \\ \equiv \underset{\theta^m_k}{\mbox{minimize}}~\frac{1}{t}\ln \bigg (\frac{1}{|\mathcal{Q}^m_k|}\sum_{\mathcal{Q}^m_k} G(\theta^m_k|\mathcal{Q}^m_k)^{-t} \bigg).\label{TERM}
\end{multline}
Here, $G(\sigma,\xi|q)= \frac{1}{\sigma}\big( 1+ \frac{\xi q}{\sigma}\big )^{-(\frac{1}{\xi}+1)}$ is the GPD's likelihood function while $t$ is the tilted factor used to leverage the effects of the outliers in model training. 
%In contrast with conventional ERM,
%
%
%
%\begin{equation}\label{MLE}
%   \underset{\sigma,\xi}{\mbox{minimize}}~ - \frac{1}{|\mathcal{Q}^m_k|}\sum_{\mathcal{Q}^m_k}\ln(G(\theta^m_k|\mathcal{Q}^m_k))
%\end{equation}
%
%
%
%in which the average of the loss function is minimized, tilted ERM considers the entropic risk measure $\frac{1}{t}\ln(\mathbb{E}_{X}[e^{tX}])$ of the loss functions as the objective which jointly incorporates the mean, variance, and other higher-order moments \cite{8936848}.
In this regard, by setting $t<0$ in \eqref{TERM}, we can account for the outliers and other extreme events.
Subsequently, based on the global model $\theta^{m-1}=\{\sigma^{m-1},\xi^{m-1}\}$ received in the $(m-1)$th interval, each sensor $k$ updates the local model parameters as per
\begin{equation}\label{para_update}
    \theta^m_k=\theta^{m-1}-\delta_{\theta}\nabla_{\theta^m_k}\tilde{\mathcal{L}}(\theta^m_k|t,\mathcal{Q}^m_k),
\end{equation}
with the initial value  $\theta^0$ and sends $\theta^m_k$ to the controller. In \eqref{para_update}, $\delta_{\theta}$ is the step size. The gradient of $\tilde{\mathcal{L}}(\theta^m_k|t,\mathcal{Q}^m_k)$ with respect to $\theta^m_k$ is given by
\begin{align}\label{gradient}
    &\nabla_{\theta^m_k}\tilde{\mathcal{L}}(\theta^m_k|t,\mathcal{Q}^m_k)= \notag
    \\&\frac{-\sum_{\mathcal{Q}^m_k} \nabla_{\theta^m_k}G(\theta^m_k|\mathcal{Q}^m_k)\times G(\theta^m_k|\mathcal{Q}^m_k)^{-t-1}}{\sum_{\mathcal{Q}^m_k} G(\theta^m_k|\mathcal{Q}^m_k)^{-t}},
\end{align}
and
\begin{align*}
  \textstyle \frac{\partial G(\theta|\mathcal{Q})}{\partial\sigma}=&\textstyle \frac{Q-\sigma}{\sigma^3}\left(1+\frac{\xi Q}{\sigma}\right)^{-2-\frac{1}{\xi}} ,
\\ \textstyle \frac{\partial G(\theta|\mathcal{Q})}{\partial\xi}=& \textstyle
  \frac{(1+\frac{\xi Q} {\sigma})^{-1-\frac{1}{\xi}}}{\sigma\xi}\left(\frac{-(\xi+1)Q}{\sigma+\xi Q}+\frac{\ln(1+\frac{\xi Q}{\sigma})}{\xi}\right).
  \end{align*}
Here, the notations $m$ and $k$ are neglected for simplicity.
The controller then calculates the global model of the $m$th updating time interval as
\begin{equation}\label{global_para}
    \theta^m = \frac{\sum^{K}_{k=1}\left (|\mathcal{Q}^m_k|\theta^m_k\right )}{\sum^{K}_{k=1}|\mathcal{Q}^m_k|}.
\end{equation}
Finally, after receiving the global feedback model $ \theta^m $, each sensor $k$ replaces the virtual queue value with
\begin{equation}\label{Upsilon_update}
    \Upsilon_k(\hat{i}+1)=\left[ \frac{\sigma^m}{1-\xi^m}-e_0\right]^+\mathbbm{E}_{ \mathcal{K}}\left[\sum^{\hat{i}}_{j=1}\mathbbm{1}_{\{q_k(j)>0\}}\right],
\end{equation}
in which $\hat{i} = \arg \min \limits_{\forall i} \{\tau_k(i)-Mm\geq0\}$,
and proceeds with the next local-model training $\theta^{m+1}_k$.
The proposed FL-based distributed power allocation is outlined in Algorithm \ref{alg}.
Therein, each sensor $k \in \mathcal{K}$ iteratively optimizes the transmission power $p_k(i_k)$ for each local-sampled data $i_k$ based on the time instance $\tau_k(i_k)$ and the FL updated virtual queue states $A(i_k),\Gamma_k(i_k),\Lambda_k(i_k),\Upsilon_k(i_k)$.

\captionsetup[algorithm]{font=footnotesize}
\begin{algorithm}[t]\footnotesize
\caption{FL-Based Distributed Power Allocation Mechanism}\label{alg}
\textbf{Input:}  $\tau_k(i_k)~\forall k\in \mathcal{K}$, \text{transmission parameters in (\ref{Tx_time})}, $\beta, \theta^0, \delta_{\theta}, M, O.$\\
\textbf{Output:} $p^*_k(i_k),A(i_k),\Gamma_k(i_k),\Lambda_k(i_k),\Upsilon_k(i_k) ~\forall i_k,~\forall k\in \mathcal{K}.$\\
\textbf{Initialize:} \text{create $i_k$ for data indicator, $i_k=1,  \forall k,$}\\ $a_k(0)=0,\eta_k(1)=0, \tau_k(i_k|i_k=0,1)=0,\forall k,\\ \{A_k(1),\Gamma_k(1),\Lambda_k(1),\Upsilon_k(1)\} = 0.$
\begin{algorithmic}[1]
\For {$m=1,2,...$} \textit{// for global model updating interval}
\For {$k=1,...,|\mathcal{K}|$}
\State {$w=1, \mathcal{Q}^m_k=\{\},$} \textit{// check each observation window}
\If {$\tau_k(i_k)\leq Mm \quad\&\&\quad w \leq W$}
\While {$\tau_k(i_k)\leq M(m-1)+Ow$}
\State \text{update} $\eta_k(i_k), \theta^1_k(i_k), \theta^2_k(i_k), \theta^3_k(i_k),$
\State \text{form $u_k(i_k)$ by \eqref{L_d_bound} and solve $p^*_k(i_k)$ by CCP.}
\State \text{count $t^*_k(i_k)$ by \eqref{Tx_time}, $E^*_k(i_k), a_k(i_k), f_k(i_k), q_k(i_k),$}
\State \text{update state $A(i_k),\Gamma_k(i_k),\Lambda_k(i_k),\Upsilon_k(i_k),$}
\State $i_k = i_k + 1,$
\EndWhile
\State \text{Count $\hat{q}^{m,w}_k$, update $\mathcal{Q}^m_k=\{\mathcal{Q}^m_k,\hat{q}^{m,w}_k\},$}
\State \text{$w=w+1,$}
\Else \text{$\mathcal{Q}^m_k=\{\},$}
\EndIf
\State \text{train $\theta^m_k$ with \eqref{para_update}, \eqref{gradient} by $\mathcal{Q}^m_k$ and update them to Controller,}
\EndFor
\State \text{Controller updates $\theta^m$ with $\theta^m_k$ by \eqref{global_para}, and broadcasts to IDs,}
\State \text{$\forall k, \Upsilon_k(i_k)$ is replaced with $\theta^m$ by \eqref{Upsilon_update}}.
\EndFor
\end{algorithmic}
\end{algorithm}

\section{Numerical Results}\label{Numerical}
We simulate a factory environment with $K=50$ sensors with $50$\,Hz data-sampling frequency in Poisson.
The considered path loss model is $32.45+31.9\log20+20\log3.5$ (dB) given $3.5$\,GHz carrier frequency and a $20$-meter sensor-controller distance \cite{nokia_spec}. 
The wireless channel experiences Rayleigh fading with unit variance.
The remaining simulation parameters are $N=3000$\,bytes, $B= 180$\,kHz, $N_0=-174$\,dBm, $p_{max}=23$\,dBm, $\beta=-2$, $f_0=5\times10^{-4}$, $e_0 = 10^{-4}$, $\epsilon=2\times10^{-3}$, $M=30$\,ms, $O=10$\,ms, $\theta^0=[0.0002,0.02]$, $\delta_{\theta}=[10^{-9},10^{-3}]$, and $t=-0.1$. 
For performance comparison, we consider four baselines: \emph{i)}
\textbf{Centralized model-training} (CENT) scheme which trains the extreme staleness GPD model at the central controller with all sensors' AoI exceedance data; 
\emph{ii)} \textbf{Local model-training} (LOCAL) scheme in which the sensors only train the GPD model individually without any information exchange;
\emph{iii)} \textbf{Non-model-training} (NonT) scheme which directly solves problem \eqref{L_form} without renewing the virtual queue $\Upsilon_k(i)$ in \eqref{Virtual_queue_2} via the training result of GPD model;
\emph{iv)} \textbf{Excess staleness-agnostic} (ESA) scheme which does not take extreme staleness into consideration, i.e., neglecting constraints \eqref{outg} and \eqref{extreme_AoI} in problem \eqref{origin}. 
In addition to the performance of the objective \eqref{main_problem}, we further investigate the expected
system energy consumption $\mathbbm{E}[E_{sys}]=\mathbbm{E}[E_k] + \mathbbm{E}[E^{train}_{comp.}]+\mathbbm{E}[E^{train}_{tx}]$, including the sensor's monitored data-updating energy $E_k$, the computation energy in GPD-model training $E^{train}_{comp.}=10^{-27} f^2_{cpu} N_{tr}l_{req}$, and the energy consumption in model-uploading $E^{train}_{tx}=p_{max}N_{tr}/\big(B\log_2\big(1+\frac{h_kp_{max}}{N_0 B}\big)\big)$.
Here, $f_{cpu}=2\times10^{11}$ cycle/s and $f_{cpu}=10^{9}$ cycle/s are the controller's and sensor's computation capabilities for model training \cite{8936848}. $N_{tr}=30$\,bytes is the single-data size in GPD-model training, and $l_{req}=87.8$ cycle/bit is the required computation frequency.
\begin{figure}[t!]
     \centering
     \begin{subfigure}[b]{0.43\textwidth}
         \centering
         \includegraphics[width=\textwidth]{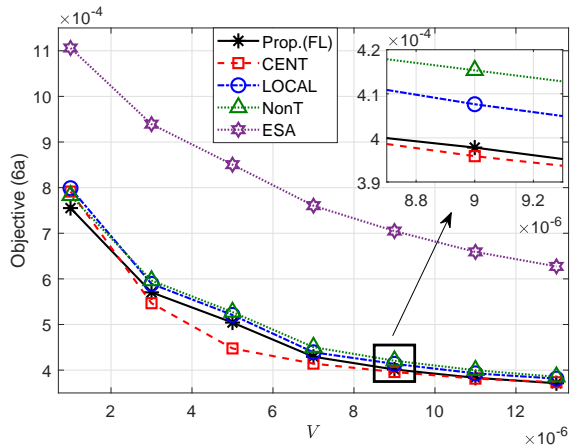}
         \caption{Value of objective \eqref{main_problem}}
         \label{fig:objective_V}
     \end{subfigure}
     \begin{subfigure}[b]{0.43\textwidth}
         \centering
         \includegraphics[width=\textwidth]{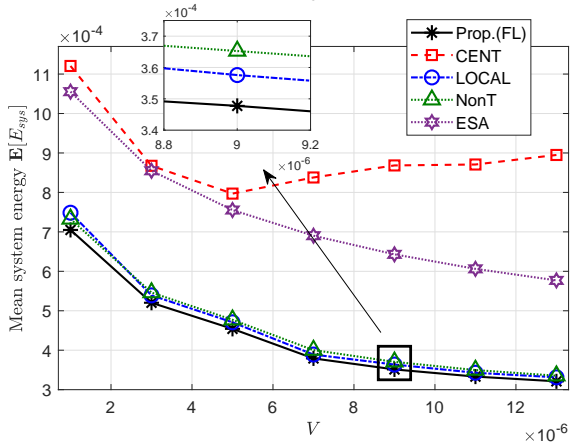}
         \caption{System energy $\mathbbm{E}[E_{sys}]$}
         \label{fig:energy_V}
     \end{subfigure}
        \caption{Energy consumption versus $V$ with $\rho=2$.}
        \label{fig:multi_V}
\end{figure}

The impact of the tradeoff parameter $V$ in the Lyapunov optimization on energy consumption is shown in Fig.~\ref{fig:multi_V}.
We first examine the performance of the objective \eqref{main_problem} in Fig.~\ref{fig:objective_V} as a function of $V$. 
Note that the objective is a decreasing function of $V$ since the importance of energy reduction increases with increasing $V$.
Although CENT outperforms the other schemes across various $V$ values due to the GPD model training, the performance of our proposed scheme is close to CENT with $0.48\%-11.26\%$ increments in the objective.
Compared with LOCAL and NonT, the proposed FL scheme decreases the objective cost up to $5.45\%$ and $5.56\%$, respectively, showing the benefit of FL model training. 
In addition, LOCAL outperforms NonT, manifesting the advantage of model training even without any information exchange.
Being agnostic to extreme staleness, ESA may incur a high instantaneous AoI, resulting in larger energy consumption. 
In this regard, ESA has a higher objective cost (up to $44.75\%$ increase) than the cost of the proposed scheme. 
\begin{figure}[t!]
     \centering
     \begin{subfigure}[b]{0.43\textwidth}
         \centering
         \includegraphics[width=\textwidth]{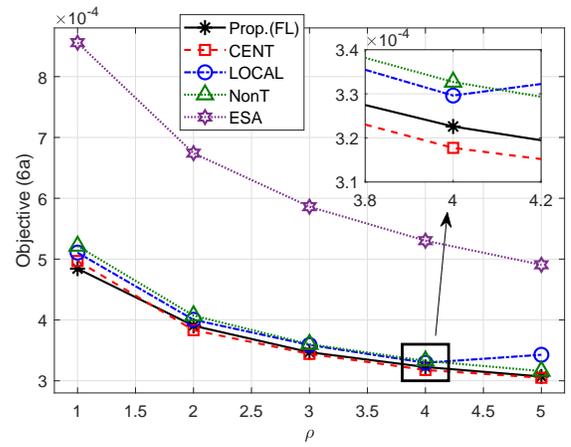}
         \caption{Value of objective  \eqref{main_problem}}
         \label{fig:objective_rho}
     \end{subfigure}
     \begin{subfigure}[b]{0.43\textwidth}
         \centering
         \includegraphics[width=\textwidth]{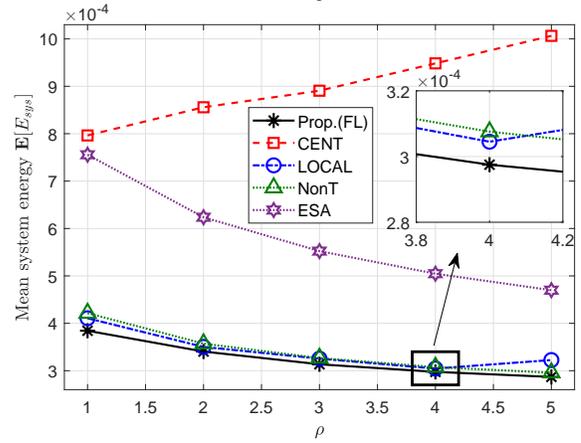}
         \caption{System energy $\mathbbm{E}[E_{sys}]$}
         \label{fig:energy_rho}
     \end{subfigure}
        \caption{Energy consumption versus $\rho$ with $V=10^{-5}$.}
        \label{fig:multi_rho}
\end{figure}

In Fig.~\ref{fig:energy_V}, we further verify the advantages of our proposed approach in terms of the expected system energy consumption.
In CENT, all sensors have to deliver every observed extreme staleness data to the controller, consuming high energy for information exchange.
Note that as $V$ increases, all schemes tend to save more power in environmental data transmission based on the  Lyapunov optimization framework but increases the occurrence chance of extreme staleness. In this situation, the CENT sensors have to consume more energy to upload more model-training data to the controller. Therefore, the expected system energy of CENT grows gradually with $V$ in Fig.~\ref{fig:energy_V}.
In the proposed FL-based approach, since the sensors train the GPD models locally and only upload the model parameters to the controller, the energy consumption is significantly reduced. 
In this regard, our approach can save up to $64.11\%$ in system energy compared to CENT. In contrast with NonT and LOCAL, our approach spends extra energy on FL model training and information exchange. 
Nevertheless, our energy-saving benefit with respect of $\mathbbm{E}[E_k]$ compensates this expenditure. 
The proposed scheme can save up to $5.82\%$ and $6.21\%$ in system energy then LOCAL and NonT. 
The objective performance and expected system energy consumption by varying $\rho$ are shown in Fig.~\ref{fig:multi_rho}.   
As per \eqref{L_form}, all the schemes put more focus on energy deduction as $\rho$ grows. 

\begin{figure}[t!]
	\begin{center}
	\includegraphics[width = .9\columnwidth]{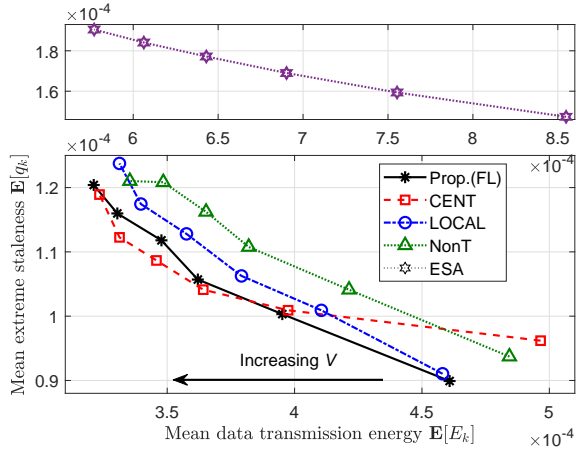}
	\caption{Extreme AoI versus average transmission energy with $\rho=2$.}
	\label{fig:Energy_vs_EAoI}
	\end{center}
\end{figure} 
\begin{figure}[t!]
	\begin{center}
	\includegraphics[width = .9\columnwidth]{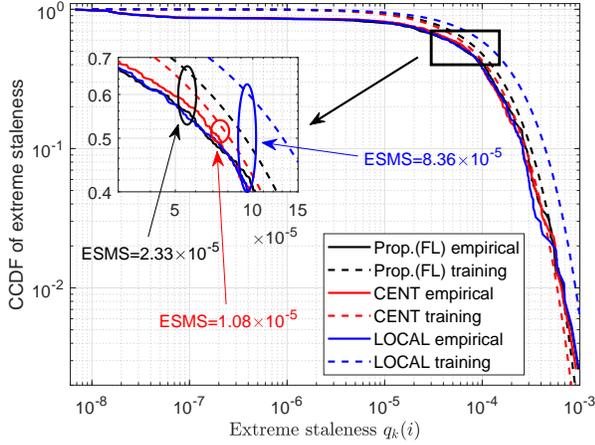}
	\caption{CCDF of extreme AoI with $\rho=2$ and $V=10^{-5}$.}
	\label{fig:extreme_GPD}
	\end{center}
\end{figure} 

The capability of extreme staleness control is manifested in Fig.~\ref{fig:Energy_vs_EAoI}. 
Given a specific amount of expected status-updating energy $\mathbbm{E}[E_k]$, the proposed approach, CENT, and LOCAL benefited from the GPD-model training showcase the lowest excess staleness values. In contrast, ESA, which is agnostic to the extreme AoI, has much higher excess staleness. However, if we further consider the energy consumption in model training, our proposed approach (compared with CENT and LOCAL) consumes less energy while achieving the same extreme staleness performance.

Finally, we discuss the performance of GPD-model training in terms of the complementary cumulative distribution function (CCDF) of extreme staleness in Fig.~\ref{fig:extreme_GPD}.
As shown, the predicted extreme staleness mean $\sigma/(1-\xi)$ is always higher than the empirical one. The reason is that the predicted extreme staleness mean values in both schemes are leveraged to suppress the extreme staleness further.  
The more precise the estimation is, the more accurate the decision.
The centralized approach, i.e., CENT, estimates the GPD model closer to the empirical one, saving more energy to control the extreme staleness.
In this regard, the proposed FL scheme and CENT, respectively, yield $2.33\times10^{-5}$ and $1.08\times10^{-5}$ in terms of
estimation-statistic mean surplus (ESMS) value between the trained model and empirical curve. 
On the other hand, the localized approach, i.e., LOCAL, has the highest ESMS $8.36\times10^{-5}$ due to the lack of global estimation.
Such results reflect on the least objective performance in Fig. \ref{fig:objective_V} and \ref{fig:objective_rho}. 
However, the mean extreme staleness from the proposed scheme is $3.54\%$ higher than CENT, underscoring that increasing transmission energy based on inaccurate predictions cannot effectively suppress extreme values.

\section{Conclusions}
We studied an industrial IoT network in which the sensors autonomously and proactively allocate transmit power for uploading monitored status data.
To avoid stale data delivery hindering the monitoring performance, we have formulated an entropic risk-minimizing problem for energy consumption subject to data staleness constraints. 
By applying extreme value theory results, we trained an GPD model to obviate extreme staleness regimes and have further proposed a distributed FL-based scheme to improve the power efficiency versus its centralized counterpart.
Numerical results have shown that the proposed FL scheme is on par with the centralized model-training scheme while consuming significantly less system energy. 
Future work will extend the current framework towards a system that can jointly optimize the data transmission power and sampling rate rather than obey a default number.

\section*{Acknowledgments}
This work is funded by the Ministry of Science and Technology (MOST) of Taiwan under the Graduate Students Study Abroad Program grant 109-2917-I-002-007 and supported by the CHIST-ERA LeadingEdge
and CONNECT projects.

\bibliographystyle{IEEEtran}
\bibliography{References.bib}
\end{document}